%% file: krad_it.tex
\documentclass[runningheads]{llncs}

\usepackage[T2A]{fontenc}
\usepackage[utf8]{inputenc}
\usepackage[english,russian]{babel}
\usepackage{graphicx}
\usepackage{color}

\usepackage{cite}
\begin{document}
\selectlanguage{english}
%
\title{Development of a Data Infrastructure\\for a Global Data and Analysis Center\\in Astroparticle Physics
\thanks{Supported by KRAD, the Karlsruhe-Russian Astroparticle Data Life Cycle Initiative (Helmholtz HRSF-0027).}}
\titlerunning{Development of a data infrastructure}


%
%
\author{Victoria Tokareva\inst{1}\orcidID{0000-0001-6699-830X}\thanks{The authors acknowledges the help of the colleagues of the
projects KCDC, KRAD, the APPDS initiative (esp. A. Kruykov, A. Mikhailov, M.D. Nguyen, A. Shigarov) and the SCC GridKa infrastructure at KIT.}, Andreas Haungs\inst{1}\orcidID{0000-0002-9638-7574}, Donghwa Kang\inst{1}\orcidID{0000-0002-5149-9767}, Dmitriy Kostunin\inst{2}\orcidID{0000-0002-0487-0076}, Frank Polgart\inst{1}\orcidID{0000-0002-9324-7146}, Doris Wochele\inst{1}\orcidID{0000-0001-6121-0632}, Jürgen Wochele\inst{1}\orcidID{0000-0003-3854-4890}}
\authorrunning{V.~Tokareva}
%
\institute{Karlsruhe Institute of Technology, Institute for Nuclear Physics, 76021 Karlsruhe, Germany\and
Deutsches Elektronen-Synchrotron, 15738 Zeuthen, Germany\\
\email{victoria.tokareva@kit.edu}}
\maketitle              
\begin{abstract}
Nowadays astroparticle physics faces a rapid data volume increase.
Meanwhile, there are still challenges of testing the theoretical models for clarifying the origin of cosmic rays by applying a multi-messenger approach, machine learning and investigation of the phenomena related to the rare statistics in detecting incoming particles.
The problems are related to the accurate data mapping and data management as well as to the distributed storage and high-performance data processing.
In particular, one could be interested in employing such solutions in study of air-showers induced by ultra-high energy cosmic and gamma rays, testing new hypotheses of hadronic interaction or cross-calibration of different experiments.
KASCADE (Karlsruhe, Germany) and TAIGA (Tunka valley, Russia) are experiments in the field of astroparticle physics, aiming at the detection of cosmic-ray air-showers, induced by the primaries in the energy range of about hundreds TeVs to hundreds PeVs.
They are located at the same latitude and have an overlap in operation runs.
These factors determine the interest in performing a joint analysis of these data.
In the German-Russian Astroparticle Data Life Cycle Initiative (GRADLCI), modern technologies of the distributed data management are being employed for establishing a reliable open access to the experimental cosmic-ray physics data collected by KASCADE and the Tunka-133 setup of TAIGA.

\end{abstract}

\section{Introduction} \input{intro.tex}

\section{Experiments}
\subsection{KASCADE experiment and data ecosystem} \input{kascade_descr.tex}

\subsection{TAIGA detector and data engineering} \input{tunka-133_decr.tex}

\section{German-Russian Astroparticle Data Life Cycle Initiative}

\input{gradlc.tex}

\subsection{KASCADE Cosmic-ray Data Center} \input{kcdc_descr.tex}

\subsection{Extending KASCADE data center with TAIGA data}
\input{appds_chall.tex}

\section{Outlook}
\input{outlook.tex}


\bibliographystyle{splncs04}
\bibliography{krad_it}

\end{document}

%% file: intro.tex

The AstroParticle Physics European Consortium (APPEC)~\cite{APPEC} considers the following challenges for the future usage of information technologies and computing in astroparticle physics: adapting the architecture of computer networks to the rapid growth of the received data amount, usage of distributed data storage and processing systems that have now found their widespread use in both industry and particle physics experiments, and problems of experimental data access and open data.

According to the Berlin Open Data Declaration~\cite{OpenAccess}, research data produced with taxpayer money must be publicly available.
Currently, the need for open access is recognized everywhere and there are several initiatives aimed at providing access to data~\cite{EOSC,ESCAPE,CERN_Open_Data}.

KASCADE~\cite{KASCADE} was one of the first experiments in the ultra-high energy field that provided access to nearly all of its data according to the principles of FAIR (Findability, Accessibility, Interoperability, and Reusability)~\cite{FAIR}.

At present, other astrophysical experiments are also moving towards publishing their data, what led to estblishing several global virtual observatories~\cite{AstroGrid,Euro-VO,SPASE}.
Following this trend, the TAIGA~\cite{Budnev:2016btu, Kostunin:2019nzy} experiment has shown interest to employ the experience gained in KASCADE for this purpose,
what led to the forming of GRADLCI~\cite{Bychkov:2018zre} aiming in developing a single center for the analysis and processing of astrophysical data with pilot datasets from both experiments KASCADE and TAIGA.
This article discusses the challenges of the data integration from various experiments, the organization of distributed access and processing, i.e.\ about the important stages of the data processing cycle, also called the data life cycle.

%% file: kascade_descr.tex


The KArlsruhe Shower Core and Array DEtector (KASCADE)~\cite{KASCADE} is an experiment in astroparticle physics that was running on Campus North of the Karlsruhe Institute of Technology (KIT) in Germany from October 1996 till December 2013, corresponding to a total of 4383 days of observation.
During this time about 450 million events were collected, which resulted in about 4~TB of reconstructed data.

The data was collected for the purpose to study the spectrum of cosmic rays in the energy range of $10^{14}$--$10^{18}$~eV.

To achieve this goal, 252 scintillation detectors were placed on the area of $200 \times 200$ m$^2$.
Later the setup was extended to KASCADE-Grande~\cite{KASCADE-Grande} and LOPES~\cite{lopes} experiments.
The high accuracy of the data collection and the large amount of accumulated statistics made it possible to obtain important results~\cite{knee,gamma-limits} in the field of ultra-high energy astroparticle physics, acknowledged by the community. 

There are several levels of reconstructed KASCADE data, starting from the original raw data stored in the CERN ZEBRA~\cite{ZEBRA} format, ending with the high-level of reconstruction shared to the general public.

\begin{figure}[htbp]\centering
  \includegraphics[width=1\textwidth]{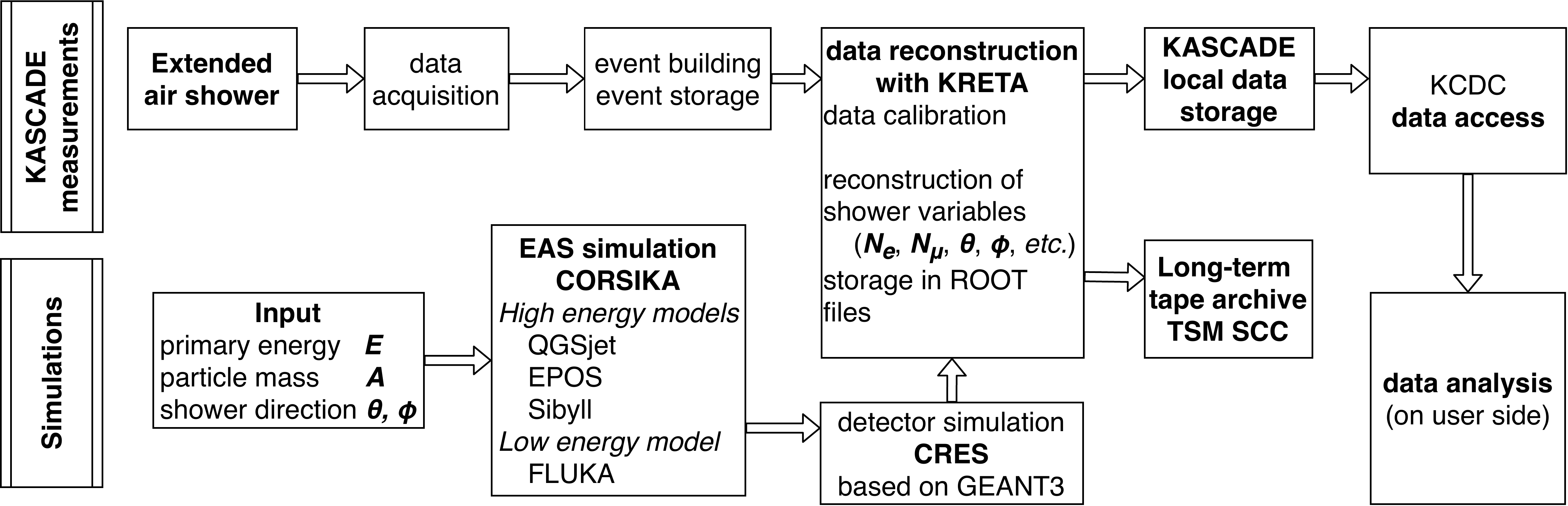}
  \caption{KASCADE data processing workflow (data life cycle).
  }
  \label{kascade_wf}
\end{figure}

Data processing is performed by means of special software developed for the experiment: a data reconstruction program KRETA~\cite{KRETA}, a program for detector output simulation CRES~\cite{CRES} based on GEANT3~\cite{geant3} and a program for detailed EAS simulation CORSIKA~\cite{CORSIKA,corsika_paper}.
A scheme of the data reconstruction process is presented in fig.~\ref{kascade_wf}.

The open access data are stored locally on KASCADE servers.
Data storage on magnetic tapes is used as a long-term storage.
It is employing the Tivoli Storage Manager (TSM) of the Steinbuch Centre for Computing (SCC) at KIT.

%% file: tunka-133_decr.tex



TAIGA (Tunka Advanced Instrument for cosmic ray physics and Gamma Astronomy) is a complex hybrid detector system, which is intended for
for cosmic ray studies from 100 TeV to several EeV as well as
for a ground-based gamma-ray astronomy for energies from a few TeV to several PeV.

The experiment infrastructure includes several setups observing air showers in a broad energy range.
They are wide-angle atmospheric Cherenkov timing arrays
Tunka-133~\cite{t133} for higher energies and TAIGA-HiSCORE~\cite{HiScore} for lower energies,
an array of imaging atmospheric Cherenkov telescopes TAIGA-IACT~\cite{IACT},
a radio extension Tunka-Rex~\cite{Bezyazeekov:2015rpa},
and a surface scintillator array Tunka-Grande~\cite{TG}.

Currently, all installations together have collected about 50 TB of raw data.
Estimates of the current annual data rate and its increase expected in the coming years are given in table~\ref{tab_taiga}.

\begin{table}[htbp]\centering
\caption{Current and expected data rates of TAIGA setups, TB/year}\label{tab_taiga}
\begin{tabular}{|l|c|c|}
\hline
Setup 				&  Current data rate 	& Expected data rate\\
\hline
TAIGA-HiSCORE			& 6.4 			&18\\
TAIGA-IACT			& 0.5			&1.5\\
Tunka-Grande, Tunka-133 and Tunka-Rex & 0.5 		&0.5\\ \hline
Total				&7.4			&20\\
\hline
\end{tabular}
\end{table}


The data collected by the experiment are stored in a distributed way on the servers of the TAIGA project in the Tunka Valley and Irkutsk, as well as on the servers of Moscow State University.
Data are stored in four specific binary data formats developed specifically for the experiment.
After being collected by different setup clusters, the events are preprocessed and merged using timestamps of the single packets.
Then the data are calibrated and stored to the server for user access.
Parsing and verifying the raw experimental data is performed using the specifications defined with FlexT and Kaitai Struct languages~\cite{Bychkov2018}.

%% file: gradlc.tex


As shown in Ref.~\cite{Apel2016}, a joint analysis of data from the certain setups of the TAIGA and KASCADE experiments is possible and of particular interest, since the experiments are at the same latitude and observe the same region of the celestial sphere, and measure the same range of the energy spectrum of cosmic rays.
Thus, a joint analysis of the data from the TAIGA and KASCADE experiments using advanced methods, including machine learning, can be significant in finding the answers to fundamental questions in astroparticle physics.
The GRADLC project was created to coordinate the joint work of two independent observatories to join efforts in building a joint data and analysis center~\cite{HaungsProc} for Multi-Messenger Astroparticle Physics.

The main goals of the project include the extension of the KCDC data center of the KASCADE experiment by adding access to the TAIGA data, software development for collaborative data analysis, providing data analysis capabilities on the data center side, and implementing solutions for visualizing analysis results.



%% file: kcdc_descr.tex
%

The KASCADE Cosmic-ray Data Center (KCDC)~\cite{kcdc-paper,KCDC} was established in 2013 to
provide users a reliable access to the cosmic-ray data collected by the KASCADE
experiment. These data include measured and reconstructed parameters of more than
433 million air showers, metadata information, simulations data for
three different high energy interaction models, published spectra from various
experiments, and detailed educational examples.
All together enable users outside the community of experts to perform their own data analysis.

With the last release, named NABOO~\cite{naboo}, more than 433 million events are provided from the whole measuring time of KASCADE-Grande.

\begin{figure}[htbp]\centering
  \includegraphics[width=0.8\textwidth]{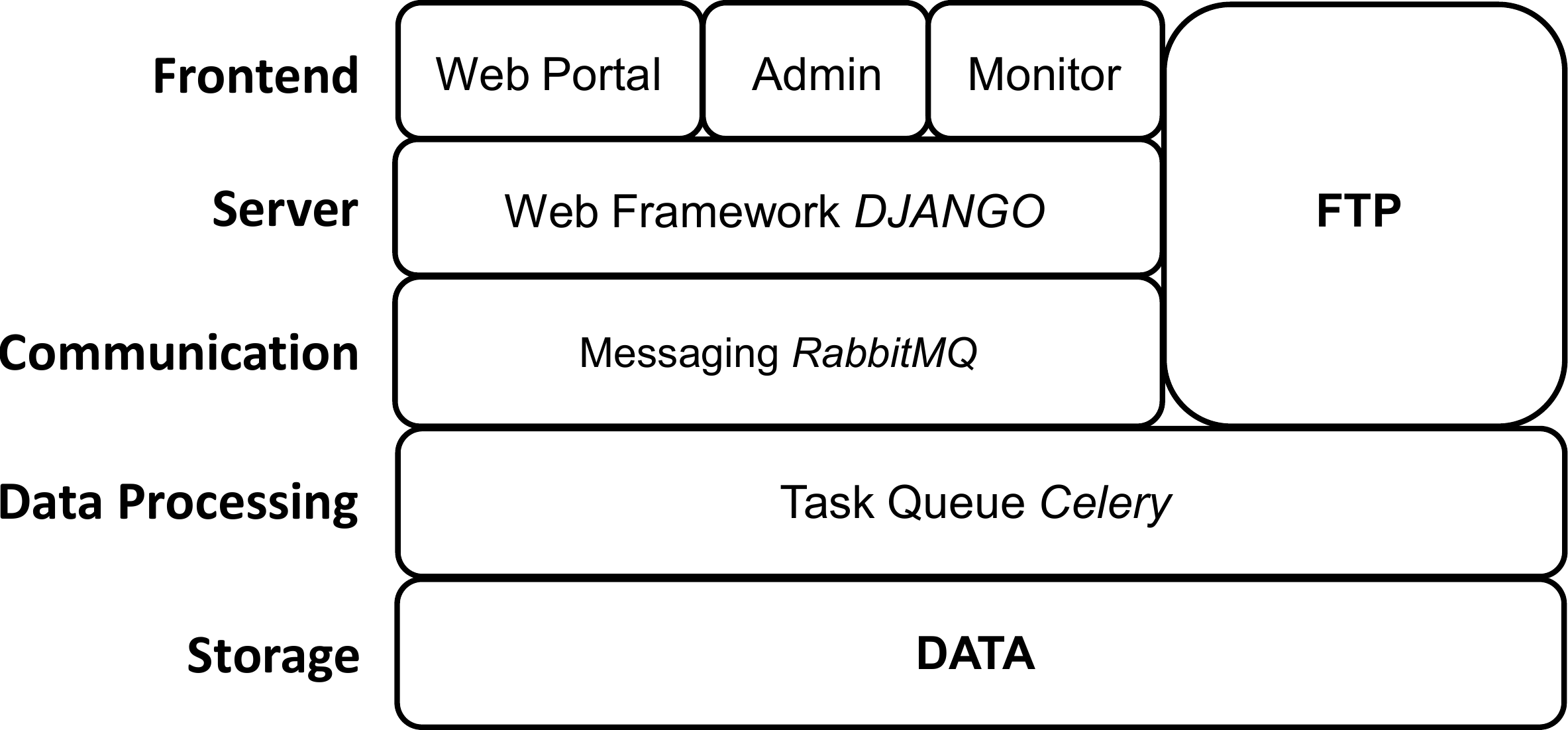}
  \caption{KCDC IT structure~\cite{kcdc-paper}.}
  \label{kcdc}
\end{figure}

%
%
The KCDC software architecture is presented in fig.~\ref{kcdc}.
Adhering to the ideas of open access, KCDC relies only on non-commercial open source software.

Expanding the experimental data by adding new detector components could require to change the structure of a stored event.
In order to do this without the restraint of a fixed database schema, a NoSQL database MongoDB has been chosen to store the experimental data.
MongoDB uses JSON-like documents with schemata.
It supports field, range query, and regular expression searches, and indexing with primary and secondary indices.
MongoDB scales horizontally using sharding and can run over multiple servers.
Currently KCDC is running MongoDB on a single server, but we are aiming at a sharded cluster for better performance.

The full KCDC system runs on an \texttt{nginx}~\cite{nginx} server and communicates with the database server and worker nodes via the \texttt{RabbitMQ}~\cite{rabbit_mq} open-source message-broker software.
The KCDC web site is built using \texttt{Django} web framework~\cite{django}, that is Python-based and follows the model-view-template architectural pattern.
Each worker node is managed and monitored via the \texttt{Celery}~\cite{celery} open source asynchronous task queue based on distributed message passing.
Python tools on the worker nodes process data selections issued by users.
The selections are stored on a dedicated FTP server, where they can be retrieved by a registered user, after the processing of their jobs has been successfully finished.

%% file: appds_chall.tex

In the process of the data center extension the following challenges appeared.

\subsubsection{Increasing data access speed.}

With the increase of the data amount, there arises a challenge of maintaining the speed of searching data on the server and providing search and selection results to the user.
At the same time, the number of requests to the data center increases, which increases the probability of a server access-denial error.
To solve this problem, a data aggregation server is introduced as an intermediate node for communication with users.
Such solution allows one to cache data which are most frequently requested for, and to reduce the server load by shielding user requests on the aggregation server, thereby helping to maintain stability and performance.

In addition, on the aggregation server one can perform a primary data search using the so-called event level metadata database.
Requests related to data analysis usually correspond to the event level, and data selections are being performed using certain criteria, such as the reconstructed energy of the event, the zenith angle, the maximum shower depth, the number of electrons, etc.

\subsubsection{Selecting database type for the metadata database.}

In modern high-load projects, NoSQL databases are in a wide use.
Due to the lack of strict requirements for a structure of the stored data, the databases of this family make it possible to easily scale the system over time, adding data of an arbitrary structure to it.
Also, these types of databases make it easy to distribute files on servers, thereby reducing the load on unified storage and facilitating data backup.
An example of such a database is MongoDB that is used to store KASCADE data.
On the other hand, SQL databases allow for very fast searches on data of fixed structure.
This is a proven database type with a well-documented standard.
The main advantage of an SQL database is declarativeness: with the help of SQL, the programmer describes only what data to extract or modify, and the exact way to do this is chosen by the database management system when processing the SQL query.
At the moment we are considering PostgreSQL as an intermediate solution: an SQL database,  which allows additional XML fields to be entered into its structure.

\subsubsection{Providing a common interface for data access.}

The KASCADE data are high-level data, while the TAIGA data are stored in a binary format.
At the same time, access should be provided at high-level of data reconstruction for all users.
To achieve this effect, we are introducing an intermediate level of data search on the TAIGA server side using file level metadata.
At the same time, events are not reconstructed at the binary level; so for the search we can use only basic information presented in the catalogs: setup, data collection season, month, day, file size, file type, etc.
The raw data found using such criteria is then transferred to the aggregation server and reconstructed there using special software for the subsequent download by the user.

%% file: outlook.tex

The GRADLC project was initiated to provide the public with access to data from two experiments, KASCADE and TAIGA.
Joint analysis of these data can bring us closer to answering fundamental questions in the field of astroparticle physics.

However, infrastructure development for data curation and joint data analysis is associated with overcoming a series of challenges, in particular, organizing a common interface for data access, expanding the current data center of KCDC while maintaining stability and performance, finding solutions for data aggregation, and some others beyond the scope of this article.

In the process of working on the project, we are trying to use proven solutions for big data processing, which are in use in the industry and particle physics. In particular, these are solutions for working with metadata, data caching and aggregation, distributed storage and processing.